\documentclass[twocolumn,preprintnumbers,amsmath,amssymb,showpacs]{revtex4}

\usepackage{color}
\usepackage{amsmath}
\usepackage{amssymb}
\usepackage{graphicx}
\usepackage{bm}
\usepackage{hyperref}
\usepackage{float}
\usepackage{mathrsfs}
\usepackage{amsfonts,dsfont} 
\usepackage{subeqnarray} 
\usepackage{array, makecell}

\def\be{\begin{eqnarray}}
\def\ee{\end{eqnarray}}

\begin{document}

\title{Thin or bulky: optimal  aspect ratios  for ship hulls}

\author{Jean-Philippe Boucher$^1$, Romain Labb\'e$^1$, Christophe Clanet$^1$ and Michael Benzaquen$^1$\footnote{Corresponding author: michael.benzaquen@polytechnique.edu}\medskip}

\affiliation{\small{ $^1$LadHyX, UMR 7646 du CNRS, \'Ecole polytechnique, 91128 Palaiseau Cedex, France}}

\date{\today}

\begin{abstract}
Empirical data reveals a broad variety of hull shapes among the different ship categories. We present a minimal theoretical approach to address the problem of ship hull optimisation. We show that optimal hull aspect ratios result -- at given load and propulsive power -- from a subtle balance between wave drag, pressure drag and skin friction. Slender hulls are more favourable in terms of wave drag and pressure drag, while bulky hulls have a smaller wetted surface for a given immersed volume, by that reducing skin friction. We confront our theoretical results to real data and discuss discrepancies in the light of hull designer constraints, such as stability or manoeuvrability. 
\end{abstract}

\maketitle

\section{Introduction}

\begin{figure}[b]
\centerline{\includegraphics[width=1\columnwidth]{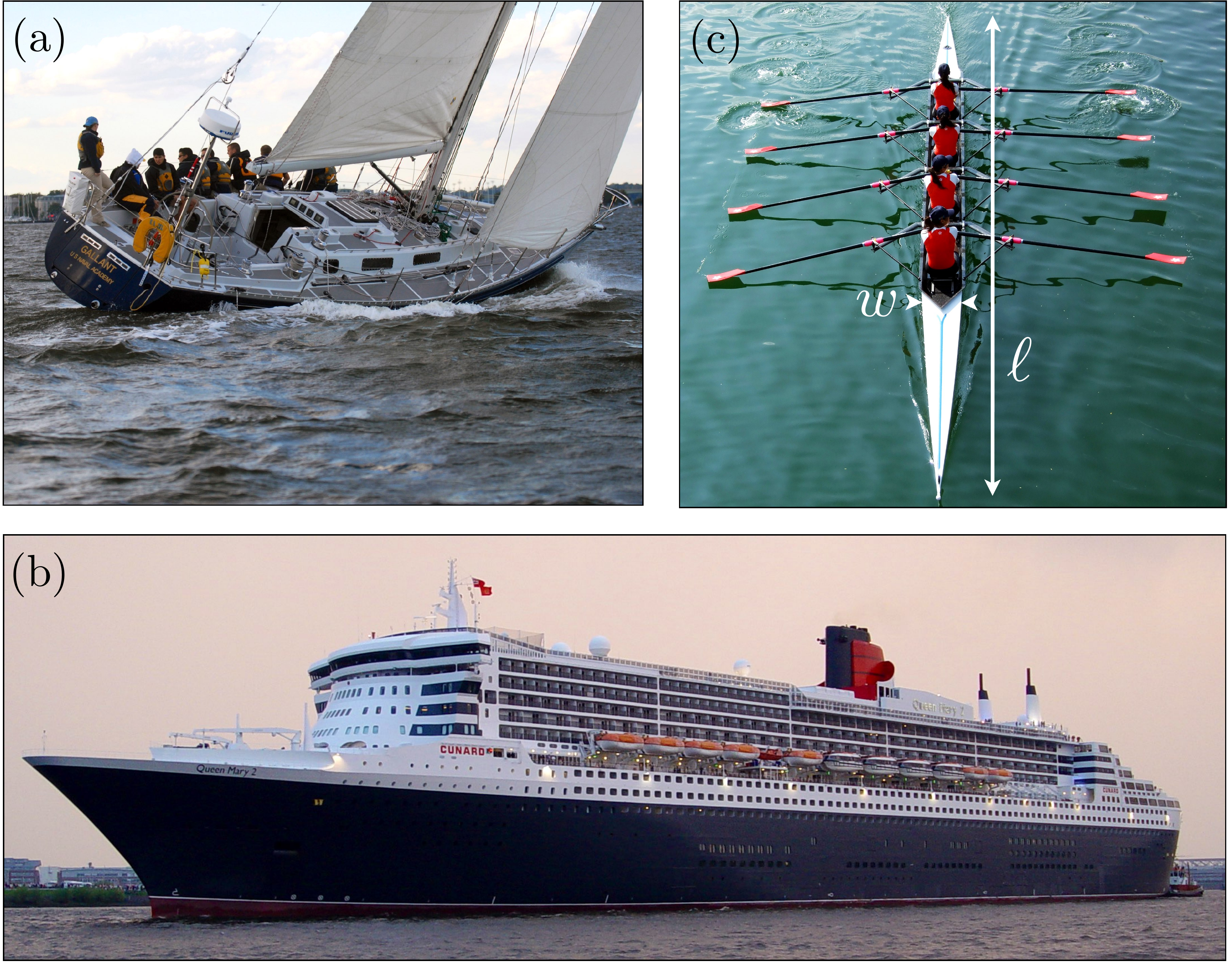}}
\caption{Pictures of (a) a 44-footer sailing boat: length-to-width aspect ratio 3 and typical Froude number 0.6,  (b)  the \emph{Queen Mary 2} liner: length-to-width aspect ratio 8.4 and typical Froude number 0.26, and (c)  a coxless quadruple scull rowing boat: length to width aspect ratio 31 and typical Froude number 0.54.  See Table \ref{table_boats} for details and characteristics of other boats.}
\label{photos_boats}
\end{figure}

The long-lived subject of ship hull design is with no doubt one of infinite complexity. Constraints may significantly vary from one ship class to another. When designing a sailing boat (see Fig.~\ref{photos_boats}(a)), stability and manoeuvrability are of paramount importance \cite{rawson2001basic,fossati2009aero,larsson2010ship,eliasson2014principles}. Liners and warships must be able to carry a maximal charge and resist rough sea conditions. Ferrys and cruising ships (see Fig.~\ref{photos_boats}(b)) must be sea-kindly such that passengers don't get sea-sick. All ship hulls share however one crucial constraint: they must suffer the weakest drag possible in order to minimise the required energy to propel themselves, or similarly maximise their velocity for a given propulsive power. 
Of particular interest is the case of rowing boats (see Fig.~\ref{photos_boats}(c)) \cite{mcarthur1997high,nolte2005rowing}, sprint canoes and sprint kayaks as they do not really have other constraints than the latter. Indeed  manoeuvrability is not relevant as they usually only have to go along straight lines, stability is at its edge and they only need to carry the athletes, {usually on} very calm waters.\\

\begin{figure}[t]
\centerline{\includegraphics[width=1\columnwidth]{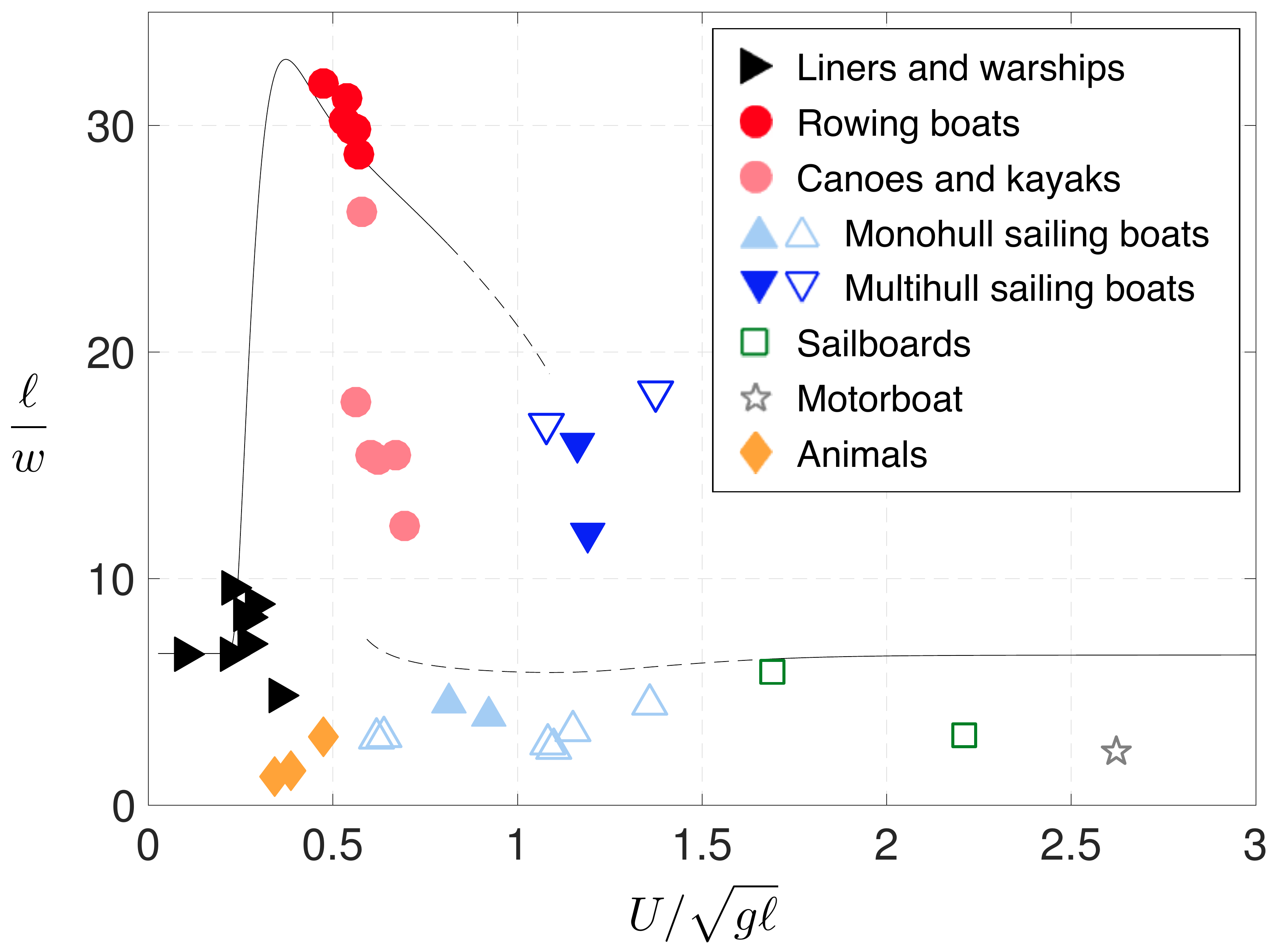}}
\caption{
 Length to width aspect ratio $\ell/w$ as a function of  Froude number $U/\sqrt{g \ell}$ for different kinds of bodies moving at the water surface (see Table \ref{table_boats} for details). Solid symbols represent displacement hulls, whereas open symbols indicate planing hulls. The aspect ratio for multihulls is computed for each hull independently.  The black line corresponds to the optimal aspect ratio, see Sect. \ref{optimal_hulls}. Solid lines indicate global optima, while dashed lines signify local optima.}
\label{aspect_ratio_real_boats}
\end{figure}

In Fig.~\ref{aspect_ratio_real_boats}, the length to width aspect ratio ($\ell/w$) of different kinds of bodies moving at the water surface is plotted against their Froude number  (see Table \ref{table_boats} for details). The Froude number is defined as $\textrm{Fr} = U/\sqrt{g \ell}$  with $U$ the hull velocity, $g$ the acceleration of  gravity and $\ell$ the length of the hull {(see Fig.~\ref{photos_boats}(c))}. As one can see, different ship categories tend to gather into clusters. These groups display very different aspect ratios, from 2-3 to about 30, even in the same Froude number regime. The highest aspect ratios are reached for rowing boats ($\ell/w \approx 30$, $\textrm{Fr} \approx 0.5$). The majority of ships stand on the left hand side of the plot ($\textrm{Fr} \lesssim 0.7$). For $\textrm{Fr} \gtrsim 0.7$, most hulls can no longer be considered as \emph{displacement} hulls (weight balanced by buoyancy) but rather as \emph{planing} hulls (weight balanced by hydrodynamic lift) and thus have a much smaller immersed volume \cite{eliasson2014principles}.  Here we  wonder how all these shapes compare to the optimal aspect ratios in terms of drag. \\

For a  fully immersed body moving at large Reynolds numbers, the drag  (also called \emph{profile drag}) is the sum of two contributions \cite{hoerner1965fluid,fossati2009aero,eliasson2014principles}: $(i)$ the skin-friction drag, which comes from the frictional forces exerted by the fluid along the surface of the body (dominant for a streamlined body, such as a plate parallel to the flow), and $(ii)$ the pressure drag, which results from the separation of the flow and the creation of vortices (dominant for a bluff body such as a sphere) \cite{hoerner1965fluid}.
One additional force arises when moving at the air-water interface: the \emph{wave resistance} or \emph{wave drag} \cite{michell1898xi,havelock1919wave,havelock1932theory}. This force results from the generation of surface waves which continuously remove energy to infinity. Thereby it is interesting to notice that many animals have air or water as a natural habitat but only a few (\emph{e.g.} ducks, muskrats or sea otters) actually spend most of their time at the water surface \cite{fish1994influence,gough2015aquatic}.\\

{As one can expect,  a number of technological advances have been developed over the years,
such as bulbous bows intended to reduce wave drag through destructive interference \cite{mccue1999bow,percival2001hydrodynamic,larsson2010ship}.} 
There exists an extended literature of numerical and experimental studies dedicated to the optimisation of ship hulls. Quite surprisingly some of them only consider wave drag in the optimisation setup (see \emph{e.g.} \cite{hsiung1981optimal,zhang2009optimization,benzaquen2014wake}). Others consider both the skin drag and the wave drag \cite{hsiung1984optimal,percival2001hydrodynamic,dambrine2016theoretical}. Very few consider the pressure drag \cite{campana2006shape} as most studies address slender streamlined bodies for which the boundary layer does not separate, leading to a negligible pressure drag. The complexity of addressing analytically this optimisation problem comes from the infiniteness of the search space. Indeed without any geometrical constraints, the functions defining the hull geometry can be anything, and computing the corresponding drag can become an impossible task.\\

Here we present a minimal approach to address the question of optimal hull aspect ratios in presence of skin drag, pressure drag and wave drag. Let us stress that we do not claim for our results to be quantitative but rather present qualitative ideas and general trends on the very complex matter of ship hull optimisation. 
We first consider a model hull shape with a minimum number of parameters and derive the expression of the total drag coefficient. Then we perform the shape optimisation at given propulsive power and load.  Finally we confront our results to the empirical data and discuss concordances and discrepancies.

\section{Wave and profile drag}

\begin{figure}[b]
\centerline{\includegraphics[width=0.8\columnwidth]{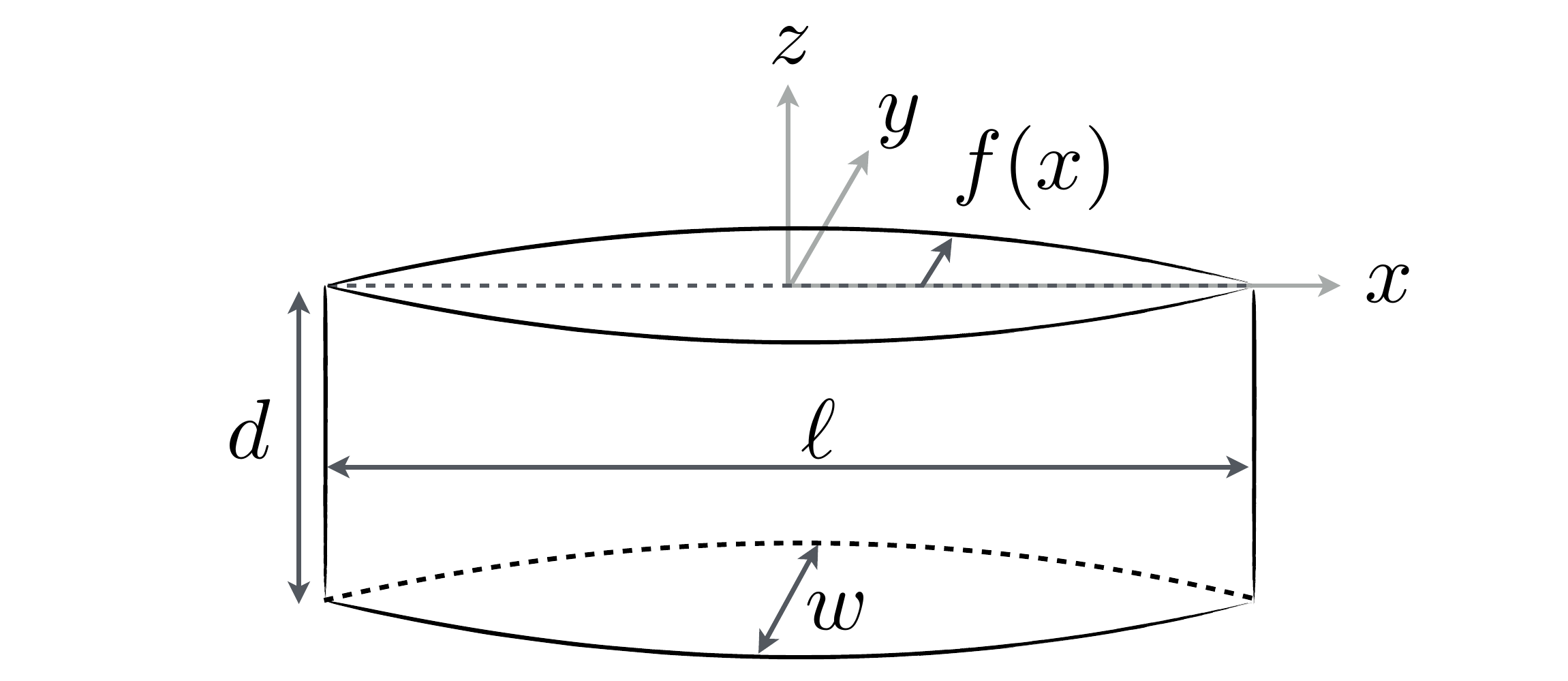}}
\caption{Schematics of the simplified hull geometry considered in this study. The hull of length $\ell$, width $w$ and draft $d$ has a constant horizontal cross-section, which is defined by $y = f(x)\mathds{1}_{z\in[-d,0]}$. Note that only the part of the hull immersed in the water is represented.}
\label{schem}
\end{figure}

In order to account in a minimal way for the wide variety of hull shapes, 
 we restrict to two-dimensional hulls (namely hulls with a constant horizontal cross-section, see Fig.~\ref{schem}).  Following the generic parametrisation of hull shapes with respect to the central plane \cite{michell1898xi,tuck1989wave,inui1962wave,wehausen1973wave}, we let $y=f(x)\mathds{1}_{z\in[-d,0]}$ the compact support hull boundary. We define the length $\ell$, width $w$ and draft $d$ and introduce the dimensionless coordinates through  ${x} = \tilde{x}{\ell}$, {${y} = \tilde{y}{\ell}$} and ${z} =\tilde{z}{\ell}$ as well as ${f}({x}) = {\tilde f( \tilde x)}{w}$ \footnote{In dimensionless coordinates, the hull boundary $y=f(x) $  with $x\in[-\ell/2,\ell/2]$ becomes $\tilde y = y/\ell = (w/\ell)\tilde f(\tilde x)$  with $\tilde x\in[-1/2,1/2]$.}. We further define the aspect ratios  $\alpha = \ell/w$  and $\beta= \ell/d$.\\

There exist two main theoretical models to estimate the wave resistance, both assuming that the fluid is incompressible, inviscid, irrotational and infinitely deep. Havelock suggested to replace the moving body by a moving pressure disturbance \cite{havelock1919wave,havelock1932theory}. This first model allows to compute the far-field wave pattern as well as  the wave resistance \cite{raphael1996capillary,benzaquen2014wake,darmon2014kelvin} but is too simple to account for the exact shape of the hull and especially to study the effect of the draft. The second model was developed by Michell for  slender bodies \cite{michell1898xi,tuck1989wave,gotman2002study}: the linearised potential flow problem with a distribution of sources on the centerplane of the hull is solved to get the expression of the wave resistance. The advantage of the latter is that it gives a very practical formula in the sense that it only takes as inputs the parametric shape of the hull and its velocity, with no need of inferring the corresponding pressure distribution.
{Using Michell's approach, we compute the wave drag}
$
R_\textrm{w} = \rho \Omega^{2/3} U^2 C_\textrm{w}
$
where $\rho$ is the water density and $\Omega = \ell w d$ \footnote{The immersed volume $\Omega_\textrm{i}$ is related to $\Omega$ through $\Omega_\textrm{i}=2a_{\tilde f}\Omega$ (see Eq.~\eqref{aftilde}).}. The wave drag coefficient $C_\textrm{w}$ writes (see Appendix A):
\be
C_\textrm{w}(\textrm{Fr}, \alpha, \beta) = \frac{4  \beta^{2/3}}{\pi  \alpha^{4/3}\textrm{Fr}^4} \, G_{\tilde f}(\textrm{Fr}, \beta) \ ,
\label{Cw_def}
\ee
where we have defined:
\be
G_{\tilde f}(\textrm{Fr}, \beta) &=& \int_{1}^{+\infty} \frac{\vert {I_{\tilde f}}(\lambda,\textrm{Fr},\beta) \vert^2}{\sqrt{\lambda^2-1}} \, \mathrm{d}\lambda \nonumber \\
{I_{\tilde f}}(\lambda,\textrm{Fr},\beta) &=& \left(1-e^{-\lambda^2/( \beta \textrm{Fr}^2)}\right)\int_{-\frac12}^{\frac12} \tilde{f}(\tilde{x}) e^{i \lambda \tilde{x}/\textrm{Fr}^2}\, \mathrm{d}\tilde{x} \ .\quad \
\label{def_G&I}
\ee
To compute the wave-drag we consider a Gaussian hull profile: 
\be
\tilde{f}(\tilde{x}) &=& \frac12 \exp[-(4\tilde{x})^2]\ . \label{parab}
\ee
This particular kind of profile allows to analytically compute the wave resistance coefficient.  The choice of this profile in comparison with more realistic profiles has no qualitative impact on our main results (see Appendix A).\\

The profile drag $R_\textrm{p}$ is the sum of the skin drag $R_\textrm{s}$  which scales with the wetted surface, and the pressure drag (or {form} drag)  $R_\textrm{f}$ which scales with the main cross-section. Given the typical Reynolds numbers for ships (ranging from $10^7$ to $10^9$), both the skin and pressure contributions scale with $U^2$ and the profile drag can be written as
$
 R_\textrm{p} =  R_\textrm{s}+ R_\textrm{f} =\rho \Omega^{2/3} U^2 C_\textrm{p}
$
with (see Appendix B):
\be
C_\textrm{p}(\alpha,\beta) = \frac{C_\textrm{d} (\alpha) \beta^{2/3}}{\alpha^{1/3}} \left[a_{\tilde f}+\frac{\alpha}{\beta} b_{\tilde f}(\alpha) \right] \ ,
\label{eq_Cs}
\ee 
where $C_\textrm{d}(\alpha)$ is the profile drag coefficient of the hull, and where:
\begin{subeqnarray}
a_{\tilde f} &=&\textstyle \int_{-\frac12}^{\frac12} \tilde{f}(\tilde{x}) \, \mathrm{d}\tilde{x} \slabel{aftilde} \\
b_{\tilde f}(\alpha) &=&  \textstyle \int_{-\frac12}^{\frac12} [{1+\tilde{f'}(\tilde{x})^2/\alpha^2}]^{1/2} \, \mathrm{d}\tilde{x} \ .
\end{subeqnarray}
The evolution of the profile drag coefficient $C_\textrm{d}$ with $\alpha$ was empirically derived for streamlined bodies \cite{hoerner1965fluid}:
$C_\textrm{d}(\alpha) = C_\textrm{f} (1+2/\alpha+60/\alpha^4)$ with $C_\textrm{f}$ the skin drag coefficient for a plate. The term $(1+2/\alpha)$ refers to the skin friction, while the term $ 60/\alpha^4$ corresponds to the pressure drag \footnote{This empirical expansion  is expected to hold for $\alpha \gtrsim 2$ (see \cite{hoerner1965fluid}).}. In the considered regimes, the skin drag coefficient is only weakly dependent on the Reynolds number \cite{hoerner1965fluid} (see Appendix B). We thus  consider here a constant skin drag coefficient $C_{\textrm f} = 0.002$, corresponding to a Reynolds number $\textrm{Re} \simeq 10^8$.\\ 

The total drag force on the hull reads $R =R_\textrm{w}+R_\textrm{p}= \rho \Omega^{2/3}U^2 C $ where $C(\alpha,\beta,\textrm{Fr}) = $
\be 
 \frac{\beta^{2/3}}{\alpha^{4/3}} \Bigg\lbrace\frac{4 }{\pi  \textrm{Fr}^4} G_{\tilde f}(\textrm{Fr}, \beta) + C_\textrm{d}(\alpha) \alpha \left[a_{\tilde f}+\frac{\alpha}{\beta} b_{\tilde f}(\alpha)\right] \hspace{-0.1cm}\Bigg\rbrace \  . \label{tot_drag}
\ee
Within the present framework and choice of dimensionless parameters, the total drag coefficient is thus completely determined by the three dimensionless variables  $\alpha$, $\beta$ and $\textrm{Fr}$, together with the function $\tilde f$.
Let us stress that this expression of the total drag coefficient is only expected to be accurate for slender hulls, as required in Michell's model \cite{tuck1989wave, gotman2002study, michelsen1960wave}.

\section{Optimal hulls} \label{optimal_hulls}

\begin{figure}[b]
\centerline{\includegraphics[width=1\columnwidth]{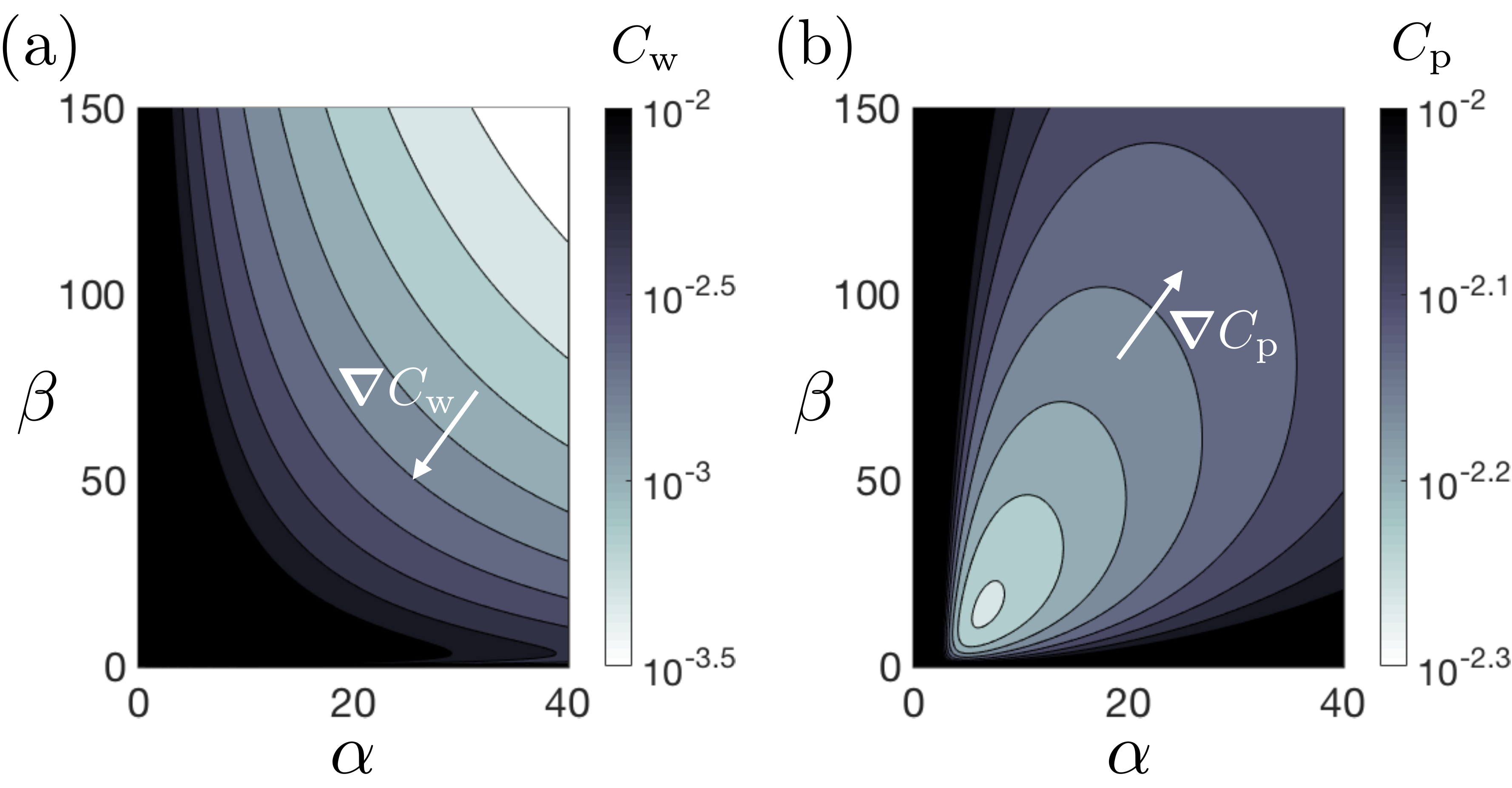}}
\caption{Contour plots of (a) the wave drag coefficient $C_\textrm{w}$ and (b) the profile drag coefficient $C_\textrm{p}$ as a function of the aspect ratios $\alpha$ and $\beta$. For the wave drag coefficient, we set $\textrm{Fr} = 0.5$. In both plots, black regions correspond to $C_\textrm{p/w} \geq10^{-2}$ and  arrows indicate the direction of the gradient.}
\label{Cs_Cp}
\end{figure}

We now seek the optimal hull shapes, that is the choice of parameters that minimises the total drag for a given load (equivalently immersed volume through Archimedes principle) and given propulsive power  -- consistent with operational conditions. Before engaging in any calculations, let us stress that the optimal aspect ratios will naturally result from a subtle balance between skin drag, pressure drag and wave drag. Indeed, on the one hand reducing skin drag amounts to minimising the wetted surface which corresponds to rather bulky hulls \footnote{
With no constraint on the geometry of the hull, the shape minimising the wetted surface is a spherical cap.}, while on the other hand reducing wave drag or pressure drag pushes towards rather slender hulls. Figure~\ref{Cs_Cp} displays the contour plots of $C_\textrm{p}$ and $C_\textrm{w}$ as function of $(\alpha,\beta)$ \footnote{Note that in order to avoid oscillations due to the sharp edges of the hull, the integral over $x$ in Eq.~\eqref{def_G&I} was actually computed over $\mathbb R$ (see Appendix A). We checked that doing so had negligible effect on the results.}. One  notices that for sufficiently large $\alpha$ and $\beta$ the gradients  $\boldsymbol \nabla C_{\textrm p}$ and $\boldsymbol \nabla C_{\textrm w}$ roughly point in opposite directions.\\

To close the problem we define the imposed propulsive power $\mathcal P = RU$. Using $U=~\textrm{Fr} [\alpha \beta \Omega g^3]^{1/6}$ one obtains:
\be
\textrm{Fr}^3 \sqrt{\alpha \beta} C(\alpha, \beta, \textrm{Fr}) = \Pi \ ,
\label{iso-puissance}
\ee
where  $C(\alpha, \beta, \textrm{Fr})$ is given by Eq.~\eqref{tot_drag}, and where we have defined the rescaled and dimensionless power:
\be
\Pi = \frac{\mathcal P}{\rho  g^{3/2}\Omega^{7/6}} \ .
\label{PiPOm}
\ee
Minimising the total drag coefficient $C$ as given by Eq.~\eqref{tot_drag} with respect to $\alpha$, $\beta$ and $\textrm{Fr}$, under the constraint given by setting the dimensionless power $\Pi$ in Eq.~\eqref{iso-puissance}, yields the optimal set of parameters ($\alpha^{\star}$, $\beta^{\star}$, $\textrm{Fr}^{\star}$) for the optimal hull geometry at given load (equivalently $\Omega$) and given propulsive power $\mathcal P$.\\

\begin{figure}[t!]
\centerline{\includegraphics[width=1\columnwidth]{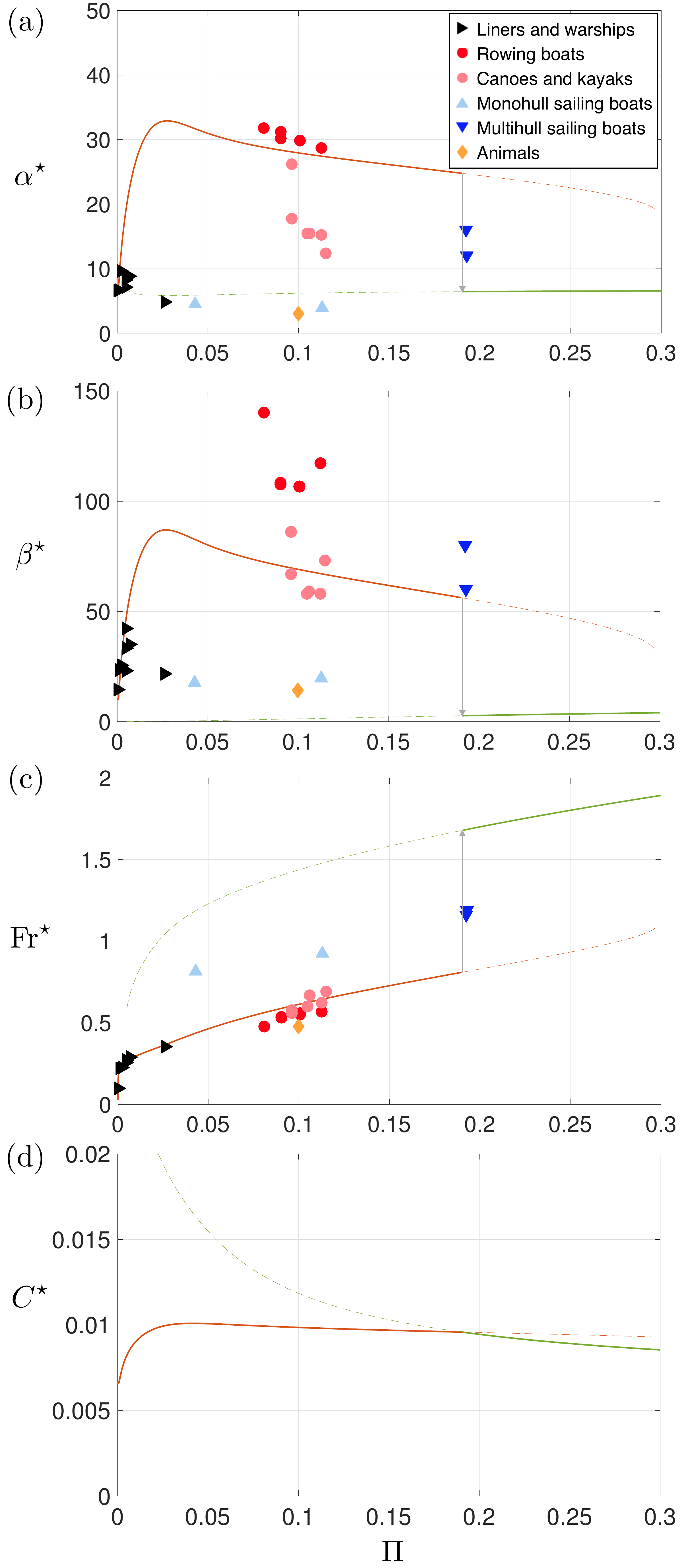}}
\caption{{ (a) Optimal aspect ratio $\alpha^{\star}$, (b) optimal aspect ratio $\beta^{\star}$, (c) optimal Froude number $\textrm{Fr}^{\star}$, and (d) corresponding value of the total drag coefficient $C^{\star} = C(\alpha^{\star}, \beta^{\star}, \textrm{Fr}^{\star})$, as a function of the dimensionless power $\Pi$. The curves in orange and green represent the two optimal branches. Solid/dashed lines indicate global/local optima.}
}
\label{optim_result}
\end{figure}

This optimisation is performed numerically using  an interior-point algorithm \cite{byrd2000trust,byrd1999interior}. The optimal parameters and the resulting total drag coefficient $C^{\star} = C(\alpha^{\star}, \beta^{\star}, \textrm{Fr}^{\star})$ as function of dimensionless power $\Pi$, are presented in Fig.~\ref{optim_result},  together with the empirical data points for comparison.
Interestingly the optimisation yields two separate solutions (see orange and green branches) corresponding to two local optima. For $\Pi \le \Pi_\textrm{c}$ (resp. $\Pi \ge \Pi_\textrm{c}$) with $\Pi_\textrm{c}\approx 0.2$, the orange (resp. green) branch constitutes the global optimum, consistent with a lower total drag coefficient $C^{\star}$ (see Fig.~\ref{optim_result}(d)). As one can see on Figs.~\ref{optim_result}(a) and (b) the optimal aspect ratios $\alpha^{\star}$ and $\beta^{\star}$ show very similar evolutions with $\Pi$.
On the one hand, both of them are maximal around $\Pi_{\max} \approx 0.03$ corresponding to $\textrm{Fr}_{\max}\approx 0.4$,
 that is the maximum wave drag regime (see Fig.~6 in Appendix A). This is consistent with the idea that thin and shallow hulls are favourable in terms of wave drag as illustrated in Fig.~\ref{Cs_Cp}(a).  
  On the other hand, for $\Pi \ll \Pi_{\max}$ or $\Pi \gg \Pi_{\max}$  the wave drag becomes negligible compared to the   profile drag, and one recovers the optimal aspect ratios in the absence of wave drag: $\alpha^\star \simeq 7$ and $\beta^\star \simeq 10$.
Figure~\ref{optim_result}(c) shows that the optimal Froude number $\textrm{Fr}^{\star}$  increases with $\Pi$. Like for $\alpha^{\star}$ and $\beta^{\star}$, there is a shift of value from $\textrm{Fr}^{\star} \approx 0.8$ to $\textrm{Fr}^{\star} \approx 1.7$, for $\Pi = \Pi_\textrm{c}$, which indicates that in this setting $0.8<\textrm{Fr}<1.7$ is never a suitable choice. This shift is also made visible in Fig.~\ref{aspect_ratio_real_boats} where the optimal aspect ratio $\alpha^{\star}$ is plotted against the Froude number.
These results obviously depend on the Reynolds number but only weakly.
Let us  stress that, while {for the optimal geometries ($\alpha^\star$, $\beta^\star$)}  the profile drag is always the dominant force regardless of the Froude number, our study shows that it is crucial to consider the wave drag in the optimisation.\\

\section{Discussion}

Our work provides a self-consistent  framework to understand and discuss the design of existing boats. Figure \ref{optim_result}  confronts the real data with the calculated optimal geometries. As one can see, while some ship categories are found in a rather good agreement with the theoretical predictions (such as liners and warships), others are very far from the computed optima (such as monohull sailing boats). Discrepancies with empirical data might primarily come from other constraints on the design of the boat which can prevail on the minimisation of the drag, such as stability, manoeuvrability, resistance to rough seas or seakindliness as mentioned in the introduction. They could also come from the assumptions of our model. In particular, a steady motion is considered here, while for rowing boats and sprint canoes, high fluctuations of speed are encountered (about 20\% of the mean velocity) and are expected to affect the total drag, notably through added mass.\\

{
{ The data for rowing boats, canoes and kayaks are found in good agreement with the optimal Froude number  $\textrm{Fr}^{\star}(\Pi)$. For rowing}
  shells, while the aspect ratios $\alpha$ are found quite close to the optimal value,
{ the aspect ratios $\beta$ lie well above the optimal curve. This indicates  that rowing shells could be shorter or have a larger draft. This discrepancy might be related to the need for sufficient spacing between rowers (long shells) and/or for stability (small draft).}
For sprint canoes and kayaks, the competition rules from the \emph{International Canoe Federation} \cite{icf_rules} impose maximal lengths for the boats \footnote{The maximal lengths for canoes and kayaks are the same for C1 and K1 (5.2 m), C2 and K2 (6.5 m) but not for C4 (9 m) and K4 (11 m) (see also Table \ref{table_boats}).} which could explain their relatively low aspect ratio $\alpha$ compared to the optimal one.  As for their aspect ratio $\beta$, contrary to rowing boats, it is found in good agreement with the optimal results.}\\

For the monohull sailing boats, the significant difference between real data and the computed optima surely comes from the need for stability (see Appendix C).  The stability of a boat mostly depends on the position of its center of gravity (which should be as low as possible) with respect to the position of the metacentre \cite{eliasson2014principles,fossati2009aero} (which should in turn be as high as possible). 
Imposing that the metacenter be above the center of gravity yields  a simple criterion for static stability \footnote{Note that for real hull design one should also address  dynamic stability \cite{rawson2001basic}, but the latter falls beyond the scope of our study.}. This is $w/d$ should be larger than a certain value  depending on  mass distribution and effective density of the hull, which  constitutes an additional constraint that could be easily taken into account in the optimisation problem.
In the simple geometry considered here and assuming a homogeneous body of density $\rho_\textrm{s}$ the latter criterion writes:  $w/d=\beta/\alpha > \psi(\rho_\textrm{s}/\rho)$ where $\psi(u) \approx 3 \sqrt{1/u-1}$ with $u \in [0, 1]$. For real boats, the critical value of $w/d$ is highly affected by the presence of a keel, intended to lower the position of the center of gravity.
 In short, stability favours wide and shallow ships. This  explains why most real data points lie below the optimal curve $\alpha^{\star}(\Pi)$ in Fig.~\ref{optim_result}(a) but above the curve $\beta^{\star}(\Pi)$ in Fig.~\ref{optim_result}(b).
Stability is all the more important for sailing boats where the action of the wind on the sail contributes with a significant destabilising torque.
Interestingly, this matter is overcome for multihull sailing boats, in which both stability and optimal aspect ratios can be achieved by setting the appropriate effective beam, namely the distance between hulls \cite{belly2004250}. This allows higher hull aspect ratios, closer to the optimal curves in Fig.~\ref{optim_result}.\\

{As displayed in Fig.~\ref{optim_result}(c), we predict a shift in the Froude number for $\Pi \approx 0.2$ which indicates that boats should not operate in the range of Froude numbers $\textrm{Fr} \in [0.8, 1.7]$.} However, when the Froude number is above $\textrm{Fr} \approx 0.7$, the hulls start riding their own bow wave: they are planing. Their weight is then mostly balanced by hydrodynamic lift rather than static buoyancy \cite{eliasson2014principles,hoerner1965fluid}. 
{ As planing is highly dependent on the hull geometry and would require to consider tilted hulls, we do not expect our model to hold in this regime. Some changes though allow to understand the basic principles.} 
 Planing drastically reduces the immersed volume of the hull which in turn reduces both the wave drag and the profile drag.
The effect on the immersed volume can be taken into account  by adding the hydrodynamic lift in the momentum balance along the vertical direction (see \footnote{
Vertical momentum balance writes $M g \simeq \rho \Omega_\textrm{i} g + \frac{1}{2}\rho C_L  \ell w \sin(2\theta) U^2$ 
 where $M$ is the mass of the boat, $\Omega_\textrm{i}$ is the immersed volume, and $\theta(\textrm{Fr})$ is the Froude-dependent angle of the hull with respect to the horizontal direction of motion \cite{hoerner1965fluid,eliasson2014principles}.
This leads to an immersed volume which depends on the Froude number through:
$\Omega \simeq ({M/\rho})/[{1+0.5 C_L \beta \sin(2\theta) \textrm{Fr}^2}] $.
For low Froude number, $\theta(\textrm{Fr}) \simeq 0$ and the volume is that imposed by static equilibrium, while for larger $\textrm{Fr}$ number $\theta > 0$ and the volume $\Omega$ is decreased. Note that {foil devices also contribute to decreasing the immersed volume, by increasing the lift.}}).\\ 

Our study  provides the guidelines of a general method for hull-shape optimisation. It does not aim at presenting quantitative results on  optimal aspect ratios, in particular due to the simplified geometry we consider and the limitations of Michell's theory for the wave drag estimation \cite{tuck1989wave, gotman2002study, michelsen1960wave}. 
Our method can be applied in a more quantitative way for each class of boat by considering more realistic hull geometries. Future work should be devoted to applying this method to the category of rowing boats, sprint canoes and sprint kayaks as these particular boats mostly require to experience the least drag, with no or little concern on stability and other constraints.

\begin{acknowledgements}
We thank Renan Cuzon, Alexandre Darmon, Fran\c{c}ois Gallaire,  Tristan Leclercq,  Pierre Lecointre, Marc Rabaud and Elie Rapha\"el for very fruitful discussions.
\end{acknowledgements}

\bibliographystyle{unsrt}
\bibliography{biblio_optimal_ratio}

\section*{Appendix}

\subsection{Wave drag coefficient}

\begin{figure}[t]
\centerline{\includegraphics[width=1\columnwidth]{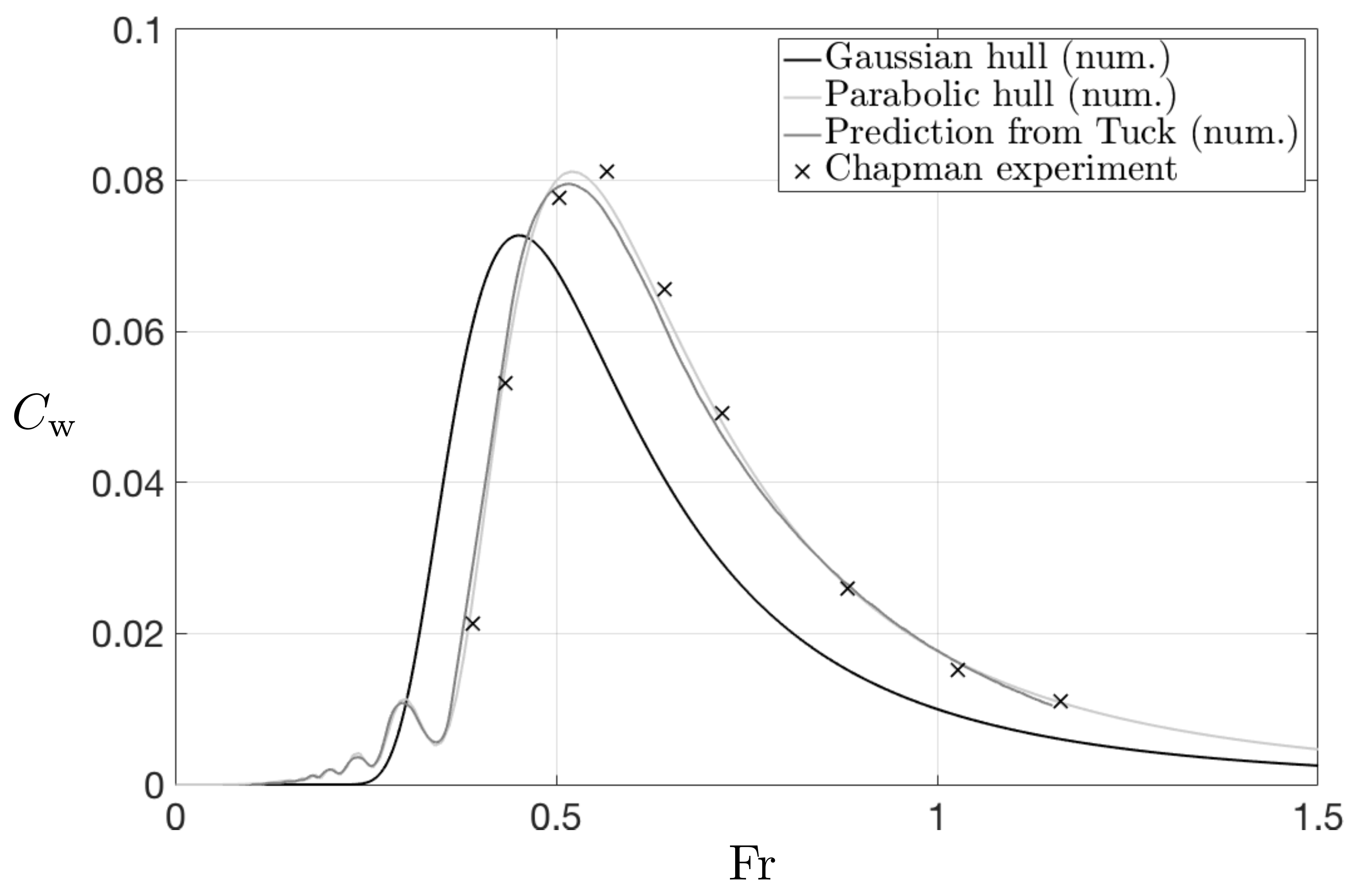}}
\caption{Wave-drag coefficient $C_\textrm{w}$ as  function of the Froude number $\textrm{Fr}$ for a Gaussian hull  and a parabolic hull for $\alpha = 6.7$ and $\beta = 2.3$. These results are compared to the theoretical curve from Tuck \cite{tuck1987wave} and experimental data points from Chapman (black crosses) \cite{chapman1972hydrodynamic}.}
\label{code_validation}
\end{figure}

Here we  derive the wave drag coefficient and discuss its behaviour for parabolic and Gaussian hull shapes. According to \cite{michell1898xi,tuck1989wave}, the wave drag in  Michell's theory writes:
\be
R_\textrm{w}(f) = \frac{4 \rho U^2}{\pi \ell^4 \textrm{Fr}^4} \int_{1}^{+\infty} \frac{\vert \mathcal I_f(\lambda,\textrm{Fr}) \vert^2}{\sqrt{\lambda^2-1}} \, \mathrm{d}\lambda \ ,
\ee
where:
\begin{equation}
\mathcal I_f(\lambda,\textrm{Fr}) = \frac{\lambda^2}{\textrm{Fr}^2} \int_{-d}^{0}\mathrm{d}z\int_{-\frac{\ell}2}^{\frac\ell2} f(x) e^{\lambda^2 z/(\ell \textrm{Fr}^2)} e^{i \lambda x/(\ell \textrm{Fr}^2)}\, \mathrm{d}x \ . \label{app_I}
\end{equation}
Taking $\Omega^{1/3}=(\ell w d)^{1/3}$ as a characteristic length, we define the wave drag coefficient through the equation $R_\textrm{w} = \rho \Omega^{2/3} U^2 C_\textrm{w}$. 
Then using the dimensionless coordinates $\tilde{x}$, $\tilde{y}$, $\tilde{z}$ and the dimensionless parameters $\textrm{Fr}$, $\alpha$, $\beta$, and integrating over $\tilde{z}$, we obtain the expression of the wave drag coefficient $C_\textrm{w}$ given in Eqs.~\eqref{Cw_def}-\eqref{def_G&I} with $I_{\tilde f} = \mathcal I_{f}/(\ell^2 w) = \mathcal I_{f}/(\Omega \beta)$.
The wave-drag coefficient compares quite well with previous numerical and experimental works \cite{tuck1987wave,chapman1972hydrodynamic} as shown in Fig.~\ref{code_validation}. This plot shows that the wave drag coefficient has the same qualitative evolution with the Froude number for a Gaussian hull and a parabolic hull. The main differences between the two are the presence of humps and hollows at low Froude number for the parabolic profile and a slight translation of the peak of wave resistance.
For a Gaussian hull, one can approximate analytically the integrals in Eq.~\eqref{app_I} by integrating $x$ over $\mathbb R$. One obtains (see Eq.~\eqref{def_G&I}):
\be
G_{\textrm{gauss}}(\textrm{Fr}, \beta) &=& \frac{\pi}{64}J\left(\frac{1}{32 \textrm{Fr}^4}\right) - \frac{\pi}{32} J\left(\frac{1}{32 \textrm{Fr}^4}+\frac{1}{\beta \textrm{Fr}^2}\right) \nonumber \\
&&+\frac{\pi}{64} J\left(\frac{1}{32 \textrm{Fr}^4}+\frac{2}{\beta \textrm{Fr}^2}\right) \ ,
\ee
where:
\be
J(u) = \int_{1}^{+\infty} \frac{e^{-u \lambda^2}}{\sqrt{\lambda^2 - 1}} \, \mathrm{d}\lambda = \frac{1}{2} e^{-u/2} \mathcal K_0(u/2) \ ,
\ee
with $\mathcal K_0(u)$  the modified Bessel function of the second kind of order zero \cite{abramowitz1964handbook}.

\subsection{Profile drag coefficient}
Here we discuss the derivation of the profile drag coefficient.
The profile drag is commonly written $R_\textrm{p} = (1/2) \rho S C_\textrm{d} U^2$ where $S$ is the wetted surface and $C_\textrm{d}$ the profile drag coefficient of the hull. Here, the wetted surface can be decomposed in two contributions $S = S_\textrm{b} + \mathcal L d$ where $S_\textrm{b} = 2 w \ell \int_{-1/2}^{1/2} \tilde{f}(\tilde{x}) \, \mathrm{d}\tilde{x}$ is the surface of the bottom horizontal cross section of the hull and $\mathcal L = 2 \ell \int_{-1/2}^{1/2} [{1+\tilde{f'}(\tilde{x})^2/\alpha^2} ]^{1/2}\, \mathrm{d}\tilde{x}$ is the perimeter of the hull. This leads to the expression of the coefficient $C_\textrm{p}$ given in Eq.~\eqref{eq_Cs}.
As mentioned in the main text, $C_\textrm{d}$ depends on the geometry through an empirical relation $C_\textrm{d}(\alpha) = C_\textrm{f} (1+2/\alpha+60/\alpha^4)$ where the skin drag coefficient $C_\textrm{f}$ weakly depends on the Reynolds number \cite{hoerner1965fluid}. In the turbulent regime  ($\textrm{Re}> 5.10^5$) one has the empirical law $C_\textrm{f}(\textrm{Re}) \simeq 0.075/(\log(\textrm{Re})-2)^2$ \cite{hadler1958coefficients}. 

\subsection{Static stability criterion}

\begin{figure}[b]
\centerline{\includegraphics[width=1\columnwidth]{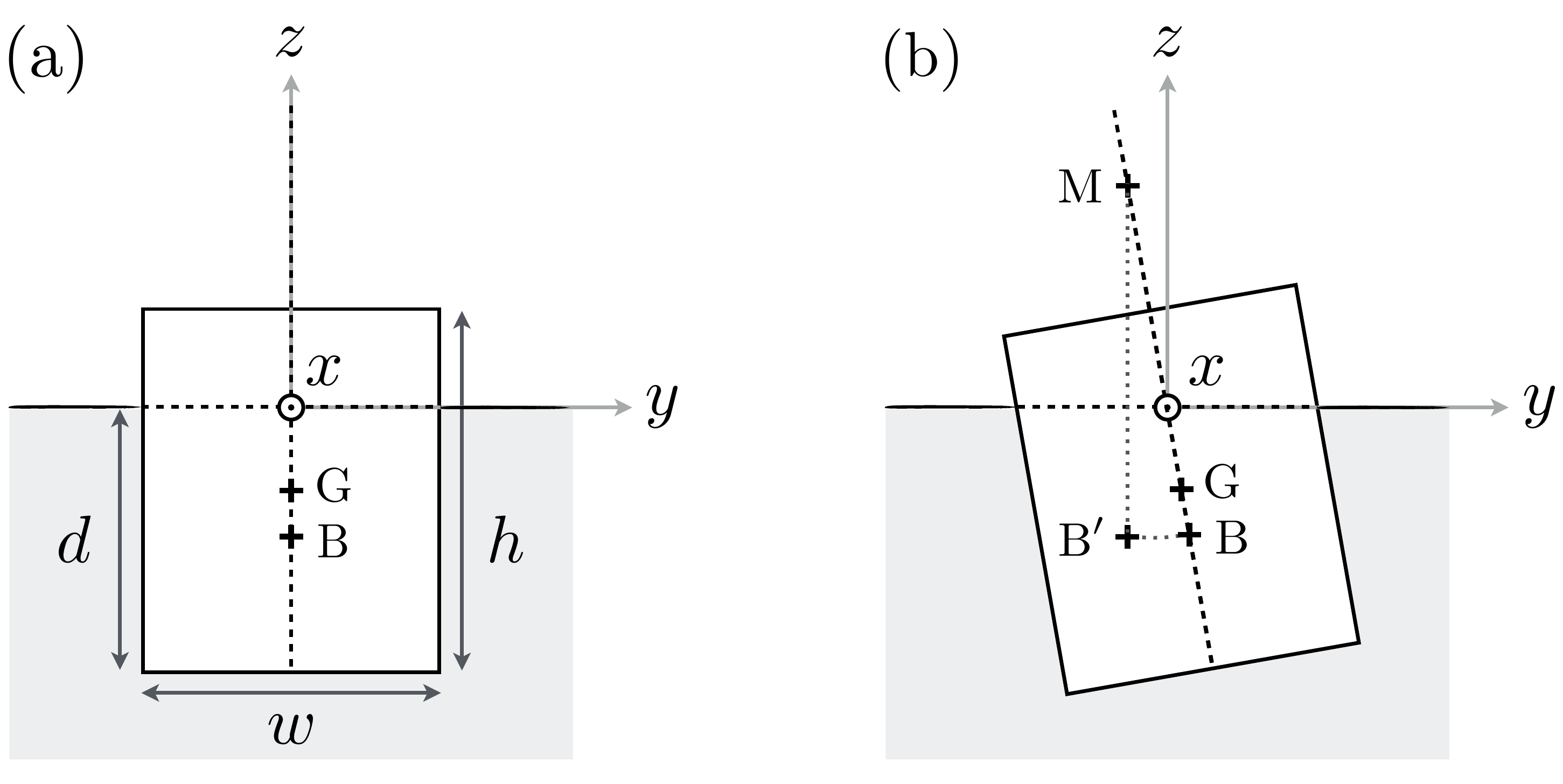}}
\caption{Cross-section of the model hull (see Fig.~\ref{schem}) in (a) vertical position and (b) slightly inclined position.}
\label{schema_stabilite}
\end{figure}

Here we  explicit the derivation of the stability criterion for the model hull presented in Fig.~\ref{schem}. Consider a homogenous body of density $\rho_s<\rho$ standing at the air-water interface (see Fig.~\ref{schema_stabilite}). We define the center of gravity G, the center of buoyancy B, and the metacenter M \cite{eliasson2014principles,fossati2009aero} as the point of intersection of the line passing through B and G and the vertical line through the new centre of buoyancy B$'$ created when the body is displaced (see Fig.~\ref{schema_stabilite}(b)). As mentioned in the main text, the stability criterion reads $\overline{\textrm{GM}}>0$, or equivalently $\overline{\textrm{BM}}>\overline{\textrm{BG}}$. On the one hand, the so-called metacentric height $\textrm{BM}$ can be computed for  small inclination angles through the longitudinal moment of inertia of the body $\mathscr I = (8c_{\tilde f}) w^3\ell/12$ with 
$
c_{\tilde f} = \textstyle \int_{-1/2}^{1/2} [ \tilde{f}(\tilde{x})]^3 \, \mathrm{d}\tilde{x}
$
and the immersed volume $\Omega_\textrm{i}= 2 a_{\tilde f}\Omega$ as:
\be
\textrm{BM} = \frac{\mathscr I}{\Omega_\textrm{i}} = \frac{c_{\tilde f}}{3 a_{\tilde f}} \frac{w^2}{d} \, .
\ee
On the other hand, one has $\textrm{BG}=(h-d)/2$ where $h$ is the total height of the hull. We then use the static equilibrium $\rho_s \Omega_\textrm{tot} = \rho \Omega_\textrm{i}$, where $\Omega_\textrm{tot} = 2 a_{\tilde f}w \ell h$ is the total volume of the body, to eliminate $h$. This finally yields the criterion $w/d > \psi(\rho_s/\rho)$ with:
\be
\psi(u) = \sqrt{\frac{3 a_{\tilde f}}{2c_{\tilde f}} \left(\frac{1}{u}-1\right)} \ , \quad u \in [0, 1]
\ee
where $\psi$ is a decreasing function of $u$. For neutrally buoyant bodies, $\psi(1) = 0$, all configurations are stable as B and G coincide. While for bodies floating well above the level of water, $\lim_{u\to 0}\psi (u)= +\infty$,  wide and shallow hulls are required to ensure stability. In the specific model case of Fig.~\ref{schem}, one has { $a_{\tilde f} \approx 0.33$, $c_{\tilde f} \approx 0.057$ and thus $\psi(u) \approx 3 \sqrt{{1}/{u}-1}$}. Taking this stability criterion into account in the optimisation procedure would reduce the search space and thus constraint the optimum curves to $\beta/\alpha > \psi(\rho_s/\rho)$.

\clearpage
\onecolumngrid
\subsection{Empirical data}
\begin{table}[h]
\centering
\begin{tabular}{|>{\raggedright\arraybackslash}l|l|r|r|r|r|r|r|}
\hline
\multicolumn{1}{|>{\centering\arraybackslash}c|}{\textbf{Category}} 
    & \multicolumn{1}{>{\centering\arraybackslash}c|}{\textbf{Boat Name}}
    & \multicolumn{1}{c|}{\thead{\textbf{Length} \\ $\ell$ (m)}}
    & \multicolumn{1}{>{\centering\arraybackslash}c|}{\thead{\textbf{Width} \\ $w$ (m)}}
    & \multicolumn{1}{>{\centering\arraybackslash}c|}{\thead{\textbf{Draft (*)} \\ $d$ (m)}}
    & \multicolumn{1}{>{\centering\arraybackslash}c|}{\thead{\textbf{Mass} \\ $M$ (kg)}} 
    & \multicolumn{1}{>{\centering\arraybackslash}c|}{\thead{\textbf{Speed} \\ $U$ (m/s)}}
    & \multicolumn{1}{>{\centering\arraybackslash}c|}{\thead{\textbf{Power (*)} \\ $P$ (kW)}}\\
\hline \hline
Liner & Titanic & 269.0 & 28.00 & 10.50 & 52300000 & 11.70 & 33833.0 \\ \hline 
Liner & Queen Mary 2 & 345.0 & 41.00 & 8.10 & 76000000 & 14.90 & 115473.0 \\ \hline 
Liner & Seawise Giant & 458.0 & 68.90 & 31.20 & 650000000 & 6.60 & 37300.0 \\ \hline 
Liner & Emma Maersk & 373.0 & 56.00 & 15.80 & 218000000 & 13.40 & 88000.0 \\ \hline 
Liner & Abeille Bourbon & 80.0 & 16.50 & 3.70 & 3200000 & 9.95 & 16000.0 \\ \hline 
Liner & France & 300.0 & 33.70 & 8.50 & 57000000 & 15.80 & 117680.0 \\ \hline 
Warship & Charles de Gaulle & 261.5 & 31.50 & 7.80 & 42500000 & 13.80 & 61046.0 \\ \hline 
Warship & Yamato & 263.0 & 36.90 & 11.40 & 73000000 & 13.80 & 110325.0 \\ \hline 
Rowing boat & Single Scull & 8.1 & 0.28 & 0.07 & 104 & 5.08 & 0.4 \\ \hline 
Rowing boat & Double Scull & 10.0 & 0.34 & 0.09 & 207 & 5.56 & 0.8 \\ \hline 
Rowing boat & Coxless Pair & 10.0 & 0.34 & 0.09 & 207 & 5.43 & 0.8 \\ \hline 
Rowing boat & Quadruple Scull & 12.8 & 0.41 & 0.12 & 412 & 6.02 & 1.6 \\ \hline 
Rowing boat & Coxless Four & 12.7 & 0.42 & 0.12 & 412 & 5.92 & 1.6 \\ \hline 
Rowing boat & Coxed Eight & 17.7 & 0.56 & 0.13 & 820 & 6.26 & 3.2 \\ \hline 
Canoe & C1 & 5.2 & 0.34 & 0.09 & 104 & 4.45 & 0.4 \\ \hline 
Canoe & C2 & 6.5 & 0.42 & 0.11 & 200 & 4.80 & 0.8 \\ \hline 
Canoe & C4 & 8.9 & 0.50 & 0.13 & 390 & 5.24 & 1.6 \\ \hline 
Kayak & K1 & 5.2 & 0.42 & 0.07 & 102 & 4.95 & 0.4 \\ \hline 
Kayak & K2 & 6.5 & 0.42 & 0.11 & 198 & 5.35 & 0.8 \\ \hline 
Kayak & K4 & 11.0 & 0.42 & 0.13 & 390 & 6.00 & 1.6 \\ \hline 
Sailing boat Monohull & Finn (p) & 4.5 & 1.51 & 0.12 & 240 & 4.10 & 4.0 \\ \hline 
Sailing boat Monohull & 505 (p) & 5.0 & 1.88 & 0.15 & 300 & 7.60 & 18.9 \\ \hline 
Sailing boat Monohull & Laser (p) & 4.2 & 1.39 & 0.10 & 130 & 4.10 & 2.7 \\ \hline 
Sailing boat Monohull & Dragon & 8.9 & 1.96 & 0.50 & 1000 & 7.60 & 16.5 \\ \hline 
Sailing boat Monohull & Star & 6.9 & 1.74 & 0.35 & 671 & 7.60 & 18.5 \\ \hline 
Sailing boat Monohull & IMOCA 60 (p) & 18.0 & 5.46 & 0.50 & 9000 & 15.30 & 843.4 \\ \hline 
Sailing boat Monohull & 18ft Skiff (p) & 8.9 & 2.00 & 0.24 & 420 & 12.70 & 85.2 \\ \hline 
Sailing boat Monohull & 49er (p) & 4.9 & 1.93 & 0.20 & 275 & 7.60 & 25.9 \\ \hline 
Sailing boat Multihull & Nacra 450 (p) & 4.6 & 0.25 & 0.12 & 330 & 9.20 & 20.7 \\ \hline 
Sailing boat Multihull & Hobie Cat 16 (p) & 5.0 & 0.30 & 0.12 & 330 & 7.60 & 20.1 \\ \hline 
Sailing boat Multihull & Macif & 30.0 & 2.50 & 0.50 & 14000 & 20.40 & 1218.3 \\ \hline 
Sailing boat Multihull & Banque populaire V & 40.0 & 2.50 & 0.50 & 14000 & 23.00 & 1701.1 \\ \hline 
Sailing boat Multihull & Groupama 3 & 31.5 & 2.40 & 0.50 & 19000 & 18.50 & 1407.3 \\ \hline 
Sailboard & Mistral One Design (p) & 3.7 & 0.63 & 0.05 & 85 & 10.20 & 6.9 \\ \hline 
Sailboard & RS:X (p) & 2.9 & 0.93 & 0.05 & 85 & 11.70 & 10.2 \\ \hline 
Motorboat & Zodiac (p) & 4.7 & 2.00 & 0.11 & 700 & 17.80 & 180.0 \\ \hline 
Animal & Swan & 0.5 & 0.40 & 0.08 & 10 & 0.76 & N.A. \\ \hline 
Animal & Duck & 0.3 & 0.20 & 0.13 & 5 & 0.66 & N.A. \\ \hline 
Animal & Human & 1.8 & 0.60 & 0.13 & 90 & 2.00 & 0.3 \\ \hline 
\end{tabular}
\caption{Characteristics of bodies moving at the water surface. The planing hulls are indicated with (p) in the column {Boat Name}. N.A. stands for \textit{Not Available}. (*) For all hulls (including planing hulls for which this estimation might be too rough), the draft is estimated using the mass of the boat and the relation $M/\rho \simeq 2 a_{\tilde f} \ell w d$ (with $a_{\tilde f}  = 0.33$). The power is estimated through diverse methods depending on the category of the boat. For liners and warships, the propulsive power can easily be found in the specification documents. For rowing boats, canoes and kayaks, we consider that the power per oarsman is 400 W. For sailing boats and sailboards, we use the sail area of the boat to derive its propulsive power (with a typical wind of 10 m/s). Note that for multihull sailing boats, the indicated dimensions correspond to one of the hulls.
}
\label{table_boats}
\end{table}

\end{document}